\documentclass[prd,reprint,showpacs,nofootinbib]{revtex4-1}
\usepackage[margin=1.82cm]{geometry}
\usepackage{array,graphicx,multirow,amsmath,amsfonts,mathrsfs}
\usepackage{nicefrac,ifthen}
\usepackage[T1]{fontenc}

\newcommand{\re}[1]{(\ref{eq:#1})}

\def\phi{\varphi}

\def\rho{\varrho}
\def\d{\mathrm{d}}
\def\p{\partial}

\renewcommand{\vec}[1]{\boldsymbol{#1}}

\newcommand{\rem}[1]{}

\def\={\discretionary{-}{-}{-}}

{\vskip 3pt plus 1pt minus 1pt\begin{normalsize}}%
{\par\end{normalsize}\vskip 3pt plus 1pt minus 1pt}

\newlength{\obrA} 
\newlength{\obrB} 

\newcommand{\fp}{\mathcal{A}}

\newcommand{\Th}{\text{\th}}
\newcommand{\Ed}{\text{\dh}}

\newcommand{\Ktau}{K_\tau}
\newcommand{\Kphi}{K_{\!\phi}}

\newcommand{\Ntau}{N_\tau}
\newcommand{\Nphi}{N_{\!\phi}}
\newcommand{\Rp}{r_{\!p\,}}
\newcommand{\Rm}{r_m}

\begin{document}

\title{Separability of test fields equations on the C\,--\,metric background II. Rotating case and the Meissner effect.}

\date{\today}
\author{David Kofro\v{n}}
\email{d.kofron@gmail.com}
\affiliation{
Institute of Theoretical Physics, Faculty of Mathematics and Physics,\\
Charles University in Prague,\\
V Hole\v{s}ovi\v{c}k\'{a}ch 2, 180\,00 Prague 8, Czech Republic}

\pacs{04.20.Jb,04.20.Cv,04.40.Nr,04.70.Bw}
\keywords{rotating C-metric, separability, Klein\,--\,Gordon equation, neutrino equation, Maxwell equation, Rarita\,--\,Schwinger equation, gravitational perturbations, Debye potentials, Meissner effect}

\begin{abstract}
We present the separation of the Teukolsky master equation for the test field of arbitrary spin on the background of the \emph{rotating} C\,--\,metric. We also summarize and simplify some known results about Debye potentials of these fields on type D background. The equation for the Debye potential is also separated.

Solving for the Debye potential of the electromagnetic field we show that on the \emph{extremely rotating} C\,--\,metric no magnetic field can penetrate through the outer black hole horizon --- we thus recover the Meissner effect for the C\,--\,metric.
\end{abstract}
\maketitle

\section{Introduction}
The pioneering work on the separability of the Teukolsky master equation on the C\,--\,metric background has been done in \cite{Prestidge1998} and generalized to the rotating case in \cite{Bini2008}. In the preceding paper \cite{KofSep1} we independently tackled the same problem and provided a separation for the ``extreme'' Newman\,--\,Penrose (NP) field components on the nonrotating C\,--\,metric background. To incorporate rotation in the C\,--\,metric is a natural generalization as some of the relativistic effects are present in rotating solutions only. As compared with \cite{Prestidge1998,Bini2008} we use a more convenient of the C\--\,metric:  (i) an explicit factorization of the structure function, which has been provided only recently in \cite{ht2}; (ii) asymptotically ``nonrotating'' form of the metric, as discussed in \cite{BiKofAcc} and (iii) the canonical coordinates; and therefore we obtain simpler results.

For the Kerr solution the magnetic and electric field becomes increasingly expelled from the horizon itself as the Kerr black hole becomes more extremal -- this is known as the Meissner effect, see e.g. \cite{Penna} for recent review or \cite{BicakJanis,DKL} for calculations on misaligned fields. This effect has been studied not only in the test field approximation, but through full exact solutions as well \cite{BicakLedvinka,KarasBudinova}.

Although the Meissner effect is known for a long time its origins have not been still explained satisfactorily. Suggestions as the infinite distance to the horizon (in the extremal case) cannot cause the Meissner effect. The infinite distance is universal property meanwhile only the axially symmetric field are subject to the Meissner effect.

The rotating C\,--\,metric is a special case of the boost\,--\,rotation symmetric solutions of Einstein (--Maxwell) field equations. At the same time it is a member of Pleba\'{n}ski\,--\,Demia\'{n}sky class of solutions \cite{pd}; therefore of algebraic type D. The C\,--\,metric represents two ``uniformly accelerated charged and rotating black holes''.

In type D spacetimes the equations for radiative (ingoing and outgoing radiation) NP field components of (a) massless Klein-Gordon field ($s=0$), (b) neutrino field ($s=\nicefrac{1}{2}$), (c) test Maxwell field ($s=1$), (d) Rarita\,--\,Schwinger field ($s=\nicefrac{3}{2}$) and (e) linear gravitational perturbations ($s=2$) can be decoupled \cite{Teuk,RS} and the Debye potential, i.e. a single scalar field from which all the field components can be generated, for them can be found \cite{Keg1,Keg2}.

We present these equations in Geroch\,--\,Held\,--\,Penrose (GHP) formalism \cite{GHP,RindlerPenrose1} which is not only succinct but provides a deeper geometrical insight than the NP formalism.

The background metric -- the rotating C\,--\,metric -- is presented in Section \ref{sec:CM}. For more details see, e.g., \cite{kw,ht2,gkp} which encompass comprehensive historical introduction. The null tetrad and corresponding spin coefficients are presented. 

In section \ref{sec:meq} the Teukolsky master equation and the equation for the Debye potential are summarized together with the separability ansatz. The equation for separated radial and angular functions are exposed.

The asymptotic behaviour of the radial function close the outer black hole horizon is analyzed in section \ref{sec:radial}. 

These general results are then used to investigate the static axially symmetric electromagnetic field in section \ref{sec:elmag}. Utilizing the Debye potential the field is easy to find. Computing the electric and magnetic flux through the horizon in the extremal case we prove that the Meissner effect works even for accelerated black holes.

\section{Charged rotating C\,--\,metric in canonical form}\label{sec:CM}
Using the signature $(-,\,+,\,+,\,+)$ the canonical form of the charged rotating C\,--\,metric is given by \cite{ht2,BiKofAcc} 
\begin{widetext}
\begin{multline}
\vec{\d} s^2 = \frac{B^2}{A^2(x-y)^2} \; \biggl\{
 \frac{\mathcal{G}(y)}{1+\left( aAxy \right)^2}\Bigl[ \left( 1+\left( aAx \right)^2 \right)\Ktau \,\vec{\d} \tau+aA\left( 1-x^2 \right)\Kphi \,\vec{\d}\phi \Bigr]^2
-\frac{1+\left( aAxy \right)^2}{\mathcal{G}(y)}\,\vec{\d} y^2 \\
+\frac{1+\left( aAxy \right)^2}{\mathcal{G}(x)}\,\vec{\d} x^2 
+\frac{\mathcal{G}(x)}{1+\left( aAxy \right)^2}\;\Bigl[ \left( 1+\left(aAy\right)^2 \right)\Kphi \,\vec{\d}\phi-aA\left( 1-y^2 \right)\Ktau \,\vec{\d} \tau \Bigr]^2\biggr\}\,,
\label{eq:rCM}
\end{multline}
\end{widetext}
where the quartic structure function $\mathcal{G}(\xi)$ can be in physically interesting cases factorized (see \cite{ht2} for detail on this factorization) as 
\begin{equation}
\mathcal{G}(\xi) = \left( 1-\xi^2 \right)\left( 1+\Rp A\xi \right)\left( 1+\Rm A\xi \right), 
\label{eq:G}
\end{equation}
where
\begin{eqnarray}
\Rp &=& m+\sqrt{m^2-q^2-a^2}\,,\\
\Rm &=& m-\sqrt{m^2-q^2-a^2}\,. 
\end{eqnarray}

The metric \re{rCM} is together with electromagnetic 4-potential 
\begin{multline}
\fp = \frac{Bqy}{1+\left( aAxy \right)^2}\Bigl[ \left( 1+\left( aAx \right)^2 \right)\Ktau \, \vec{\d} \tau \\ + aA \left(1-x^2  \right)\Kphi \,\vec{\d}\phi \Bigr] ,
\label{eq:r4p}
\end{multline}
a solution of the Einstein\,--\,Maxwell equations for $\tau\in \mathbb{R}$, $x\in \mathbb{R}$, $y\in\mathbb{R}$ and $\phi \in \langle 0, \,2\pi \rangle$ in general.

We are interested in spacetime which can be interpreted as uniformly accelerated black hole with acceleration $A>0$. This can be achieved by defining the ranges of values $\Rp$ and $\Rm$, which are roots of the structure function $\mathcal{G}(\xi)$ rescaled by $A$ by 
\begin{equation}
-\infty \leq -1/A\Rm \leq -1/A\Rp < -1 \,.
\end{equation}
Also the admissible values of coordinates are restricted to
\begin{align}
x & \in \langle -1,\, 1 \rangle\,, &
y & \in \langle -1/A\Rm,\, 1 \rangle\,, &
x-y & > 0\,,
\label{eq:range}
\end{align}
where $y=-1/A\Rm$ is the inner black hole horizon, $y=-1/A\Rp$ is the outer black hole horizon, $y=-1$ is acceleration horizon and $x-y=0$ is asymptotic infinity. In these coordinates the axis is divided in two parts by the black hole: $x=1$ corresponds to the part of axis connecting black hole horizon with the acceleration horizon and the part of axis where $x=-1$ connects black hole horizon with infinity. 

The coordinate singularity at the acceleration horizon in \re{rCM} can be avoided in the global boost\,--\,rotation symmetric coordinates \cite{gkp}. Only in these global coordinates the second black hole appears after we proceed with the analytic continuation across the acceleration horizon.

The basic set of parameters, which enters the metric, is the acceleration $A$,the rotation $a$, and the position of outer $\Rp$, resp. inner $\Rm$, horizon. $\Rm$. But the mass and the charge parameters are encoded in the metric via relations $m=(\Rp+\Rm)/2$ and $q^2=\Rp\Rm-a^2$, as follows from \re{G}. Only three of these parameters are independent. 

From the newly introduced dimensionless constants $\Kphi$, $\Ktau$ and $B$ only the $\Ktau$ explicitly demonstrates coordinate freedom. The other ones, $\Kphi$ and $B$, change the physical properties of the solution. The constant $B$ serves as a conformal factor which cannot be transformed out by a coordinate transformation and $\Kphi$ defines the conical singularities along the axis. We keep, however, even the $\Ktau$, because it is useful in demonstrating the ``visual symmetry'' of some equations.

For the sake of simplicity of our relations, let us also define function $\mathcal{H}(\xi)$, conformal factor $\Omega_0$ and constant $\Gamma$ as
\begin{align}
\mathcal{H}(\xi) &\equiv \frac{\mathcal{G}(\xi)}{1+\left( aAxy \right)^2}\,,&
\Omega_0 &= \frac{B}{A(x-y)}\,,&
\Gamma &= \sqrt{1+a^2\!A^2}\,.
\label{eq:H}
\end{align}

Moreover, we have to introduce a null tetrad $(\vec{l},\,\vec{n},\,\vec{m},\,\bar{\vec{m}})$ such that $l^an_a = -1$ and $m^a\bar{m}_a=1$ as follows
\begin{widetext}
\begin{align}
\vec{l} &= -\frac{\Omega_0}{\sqrt{2}}\left[ 
  \left( 1+a^2\!A^2x^2 \right)\sqrt{-\varepsilon\mathcal{H}(y)}\, \Ktau \,\vec{\d}\tau 
  - \frac{1}{\sqrt{-\varepsilon\mathcal{H}(y)}}\;\vec{\d} y 
  + aA\left( 1-x^2 \right)\sqrt{-\varepsilon\mathcal{H}(y)}\,\Kphi \,\vec{\d}\phi \right],  \label{eq:PND-l}\\ 
\vec{n} &= -\frac{\Omega_0}{\sqrt{2}}\left[ 
  \left( 1+a^2\!A^2x^2 \right)\sqrt{-\varepsilon\mathcal{H}(y)}\, \Ktau \,\vec{\d}\tau 
  + \frac{1}{\sqrt{-\varepsilon\mathcal{H}(y)}}\;\vec{\d} y 
  + aA\left( 1-x^2 \right)\sqrt{-\varepsilon\mathcal{H}(y)}\,\Kphi \,\vec{\d}\phi \right], 
  \label{eq:PND-n} \\
\vec{m} &= 
  \frac{\Omega_0}{\sqrt{2}}\, \left[ 
  - iaA\left( 1-y^2 \right)\sqrt{\mathcal{H}(x)}\,\Ktau \,\vec{\d}\tau 
  - \frac{1}{\sqrt{\mathcal{H}(x)}}\;\vec{\d} x 
  + i\left( 1+a^2\!A^2y^2 \right)\sqrt{\mathcal{H}(x)}\, \Kphi \,\vec{\d}\phi \right], \label{eq:PND-m} 
\end{align}
\end{widetext}
and $\vec{\bar{m}}$ is given as a complex conjugate of $\vec{m}$. This null tetrad definition is valid in the region between outer black hole horizon and acceleration horizon, where $\mathcal{G}(y)<0$ for $\varepsilon=1$. In the regions where $G(y)>0$ the parameter $\varepsilon$ has to be set to -1. Note that this does not affect the separability of the equation at all.

In this tetrad $\vec{l}$ and $\vec{n}$ are parallel to principal null directions, but the vector field itself are not tangent to affinely parameterized null congruences.

There exist two Killing vector fields
\begin{align}
\vec{\xi_{\tau}} &= \Ntau \, \frac{\vec{\p}}{\vec{\p \tau}}\,, &
\vec{\xi_\phi} &= \Nphi \,\frac{\vec{\p}}{\vec{\p\phi}}\,,
\label{eq:KV}
\end{align}
although the normalization of axial Killing vector field $\vec{\xi_\phi}$ is quite obvious, this is not true for boost Killing vector field $\vec{\xi_\tau}$. Therefore, we leave both these constants in subsequent equations.

According to \cite{KrKuCKYPD}, there are also nontrivial conformal Killing\,--\,Yano tensors which we readapted to our metric and coordinates:
\begin{eqnarray}
\vec{k} &=& \phantom{-}\Omega_0\,\Big[\, aAxy\, \vec{l} \wedge \vec{n} -i \vec{m}\wedge\vec{\bar{m}}\, \Big]\,, \\
\star\vec{k} &=& -\Omega_0 \, \Big[\, \vec{l}\wedge \vec{n}+iaAxy\,\vec{m}\wedge \bar{\vec{m}}\, \Big]\,.
\label{eq:cky}
\end{eqnarray}
This directly leads to a conformal Killing tensor $\hat{Q}_{ab} = k_{ac}k^{c}_{\phantom{c}b} = \frac{1}{2}\left[ \left( aAxy \right)^2\,\vec{l}\otimes\vec{n}+\vec{m}\otimes\bar{\vec{m}} \right].$
 
The NP spin coefficients corresponding to the tetrad \re{PND-l}\,--\,\re{PND-m} are
\begin{align}
\pi & = \phantom{-}\frac{1}{\sqrt{2}}\,\frac{A}{B}\,\frac{\left( 1+iaAxy \right)\left( 1-iaAy^2 \right)\sqrt{\mathcal{G}(x)}}{\left( 1+(aAxy)^2 \right)^{\nicefrac{3}{2}}}\,, \\
\mu &= -\frac{1}{\sqrt{2}}\,\frac{A}{B}\,\frac{\left( 1+iaAxy \right)\left( 1-iaAx^2 \right)\sqrt{-\varepsilon\mathcal{G}(y)}}{\left( 1+(aAxy)^2 \right)^{\nicefrac{3}{2}}}\,, \\
\epsilon &= \frac{1}{4\sqrt{2}}\,\frac{A}{B}\, \frac{\left( 1-iaAxy \right)^2\left( x-y \right)^3}{\sqrt{1+(aAxy)^2}\sqrt{-\varepsilon\mathcal{G}(y)}} \nonumber\\
 &\qquad\qquad\qquad\times \frac{\p}{\p y} \left( \frac{-\varepsilon\mathcal{G}(y)}{\left( 1-iaAxy \right)^2\left( x-y \right)^2} \right)\,, \\
\beta &= -\frac{1}{4\sqrt{2}}\,\frac{A}{B}\, \frac{\left( 1-iaAxy \right)^2\left( x-y \right)^3}{\sqrt{1+(aAxy)^2}\sqrt{\mathcal{G}(x)}} \nonumber\\
 &\qquad\qquad\qquad\times \frac{\p}{\p x} \left( \frac{\mathcal{G}(x)}{\left( 1-iaAxy \right)^2\left( x-y \right)^2} \right)\,,
\label{eq:sc}
\end{align}
and then 
\begin{align}
\pi &= -\tau\,, & \rho &= \varepsilon\mu\,, & \epsilon &= \varepsilon\gamma\,, & \beta &= -\alpha\,, \\
\kappa &=0\,, & \sigma &= 0\,, & \nu &= 0\,, & \lambda &= 0\,.
\label{}
\end{align}
Finally, the only non-zero Weyl NP scalar is
\begin{align}
\psi_2 &=  \frac{1}{12}\,\frac{A^2\left( x-y \right)^2}{B^2} \, \frac{\left( 1-iaAxy \right)^3}{\left( 1+(aAxy)^2 \right)}\times\\
&\left[ \frac{\p^2}{\p x^2}\left( \frac{\mathcal{G}(x)}{\left( 1-iaAxy \right)^3} \right) - \frac{\p^2}{\p y^2}\left( \frac{\mathcal{G}(y)}{\left( 1-iaAxy \right)^3} \right)\right]. \nonumber 
\end{align}

\section{Master equation}\label{sec:meq}
Teukolsky \cite{Teuk} provided decoupled equations for NP components of gravitational perturbation $\Psi_0$ and $\Psi_4$, of test electromagnetic field $\Phi_0$ and $\Phi_2$ and neutrino field $\chi_0$ and $\chi_1$ in general type D spacetime. For the Rarita\,--\,Schwinger field this was done in \cite{RS}. In \cite{KofSep1} we summarized these equations using GHP formalism. However, we realized that it is not sufficient in general to denote the NP field components by its spin weight $s$ only, but it is necessary to take the spin $S$ of the field into account as well. Here we, therefore, rewrite the equations, using a little bit different notation. Namely, every field component is denoted as $\Phi_{[p,q]}^{(s,S)}$ where $p$ and $q$ are GHP weights, $s$ is the spin weight\footnote{Of course, $s=\nicefrac{1}{2}(p-q)$.} and $S$ is spin of the field, see Tab. \ref{tab:T1}.
\begin{table*}
\centering
\caption{Notation: spin a GHP weight of field components.}
\renewcommand{\arraystretch}{1.5}
\begin{ruledtabular}
\begin{tabular}{l|ccccccccc} 
 & 
$\Psi_4$ & $\Sigma^{RS}_3$ &
$\Phi^{EM}_2$ & $\chi_1$ &
$\Phi^{KG}$ &
$\chi_0$ & $\Phi^{EM}_0$ &
$\Sigma^{RS}_0$ & $\Psi_0$ \\
$\Phi$ & 
$\Phi^{(-2,2)}_{[-4,0]}$ & $\Phi^{(-\nicefrac{3}{2},\nicefrac{3}{2})}_{[-3,0]}$ &
$\Phi^{(-1,1)}_{[-2,0]}$ & $\Phi^{(-\nicefrac{1}{2},\nicefrac{1}{2})}_{[-1,0]}$ &
$ \Phi^{(0,0)}_{[0,0]}$ & 
$\Phi^{(\nicefrac{1}{2},\nicefrac{1}{2})}_{[1,0]}$ &$\Phi^{(1,1)}_{[2,0]}$& 
$\Phi^{(\nicefrac{3}{2},\nicefrac{3}{2})}_{[2,0]}$ &$\Phi^{(2,2)}_{[4,0]}$  
\end{tabular}
\end{ruledtabular}
\label{tab:T1}
\end{table*}

In general the equation for $\Phi^{(S,S)}_{[2S,0]}$ field component reads
\begin{multline}
\bigl[ \left( \Th -\bar{\rho}-2S\rho \right)\left( \Th' -\rho' \right) - \left( \Ed-\bar{\tau}'-2S\tau \right)\left( \Ed'-\tau' \right)\\ -(2S-1)(S-1)\psi_2 \bigr] \Phi^{(S,S)}_{[2S,0]}  = 0 \,, 
\label{eq:SWplus}
\end{multline}
with the separable ansatz\footnote{The function $\mathcal{X}$ should bear also indices $l,\,m,\,\omega$ as follows from the necessity to label the basis of the solutions of Sturm\,--\,Liouville problem. And so does $\mathcal{Y}$. We omit these indices -- the reader is kindly asked to imagine them anywhere $\mathcal{X}$ and $\mathcal{Y}$ is present.}
\begin{multline}
\hat{\Phi}^{(S,S)}_{[2S,0]} = e^{-i\omega\tau} e^{im\phi} \left( x-y \right)^{1+S} \\
  \times \left( 1-iaAxy \right)^{-S} 
  \mathcal{X}^{(S)}_{(S)}(x)\,\mathcal{Y}^{(-S)}_{(S)}(y)\,. \label{eq:sepSWp} 
\end{multline}
For field component $\Phi^{(-S,S)}_{[-2S,0]}$ the equation is
\begin{multline}
\bigl[ \left( \Th'-\bar{\rho}'-2S\rho' \right)\left( \Th-\rho \right)-\left( \Ed'-\bar{\tau}-2S\tau' \right)\left( \Ed-\tau \right)\\ -\left( 2S-1 \right)\left( S-1 \right)\psi_2 \bigr] \Phi^{(-S,S)}_{[-2S,0]} = 0\,,
\label{eq:SWminus}
\end{multline}
and the separable ansatz reads
\begin{multline}
\hat{\Phi}^{(-S,S)}_{[-2S,0]} = e^{-i\omega\tau} e^{im\phi} \left( x-y \right)^{1+S}\\
  \times \left( 1-iaAxy \right)^{-S}  
  \mathcal{X}^{(-S)}_{(S)}(x)\,\mathcal{Y}^{(S)}_{(S)}(y)\,. \label{eq:sepSWm}
\end{multline}
Observe that Eq. \re{SWminus} is just a primed version of Eq. \re{SWplus}.

In \cite{Keg1} and \cite{Keg2} Cohen and Kegeles provided Debye potentials for field of arbitrary spin (the discussion of $S=\nicefrac{3}{2}$ is done in \cite{RS2}). In a slightly modified notation ($s\rightarrow -S$ and complex conjugation), and rewritten in GHP formalism, this equation reads 
\begin{multline}
\bigl[ \left( \Th'-\rho' \right)\left( \Th+(2S-1)\bar{\rho} \right)-\left( \Ed-\tau \right)\left( \Ed'+(2S-1)\bar{\tau} \right) \\
-\left( 2S-1 \right)\left( S-1 \right)\bar{\psi_2} \bigr] \hat{\psi}_{[0,-2S]}\equiv \square_S\hat{\psi}_{[0,-2S]} = 0\,.
\label{eq:DebP}
\end{multline}
which admits a separable ansatz
\begin{multline}
\hat{\psi}_{[0,-2S]} = e^{-i\omega\tau} e^{im\phi} \left( x-y \right)^{-S+1}  \\
 \times \left( 1+iaAxy \right)^S \mathcal{Y}^{(S)}_{(S)}(y)\mathcal{X}^{(S)}_{(S)}(x) \,.\label{eq:sepDeb}
\end{multline}
The operator $\square_S$ defined in Eq. \re{DebP} is invariant under the composition of $^*$ and $'$ operation, i.e. $\square_S=\square_S^{\prime *}=\square_S^{*\prime}$. From this follows $\hat{\psi}_{[0,-2S]} = \pm (\hat{\psi}^{\prime *})_{[0,-2S]} = \pm (\hat{\psi}^{*\prime})_{[0,-2S]}$.

The equations \re{SWplus}, \re{SWminus} and \re{DebP} with ansatz \re{sepSWp}, \re{sepSWm} and \re{sepDeb} leads to separated equations for $\mathcal{X}^{(s)}_{(S)}$ and $\mathcal{Y}^{(s)}_{(S)}$ where $s=\pm S$ and the indices $l,\,m$ and $\omega$ are again omitted
\begin{widetext}
\begin{align}
\frac{\left[ \mathcal{G} \mathcal{X}^{(s)}_{(S),x} \right]_{,x}}{{\mathcal{X}^{(s)}_{(S)}}} +  \frac{s^2\!+\frac{1}{2}}{3} \; \mathcal{G}_{,xx} + \Lambda^{(s)}_{(lm)} - \frac{\left( \left( 1+a^2\!A^2x^2 \right)\hat{m} -\frac{s}{2}\,\mathcal{G}_{,x} + aA\left( 1-x^2 \right)\hat{\omega} \right)^2}{\mathcal{G}} - 4aA\left( aA\hat{m}-\hat{\omega} \right)sx  &= 0\,, \label{eq:gX} \\
\frac{\left[ \mathcal{G} \mathcal{Y}^{(s)}_{(lm),y} \right]_{,y}}{{\mathcal{Y}}^{(s)}_{(S)}} +  \frac{s^2\!+\frac{1}{2}}{3} \; \mathcal{G}_{,yy} + \Lambda^{(s)}_{(S)} - \frac{\left(i \left( 1+a^2\!A^2y^2 \right)\hat{\omega} +\frac{s}{2}\,\mathcal{G}_{,y} -i aA\left( 1-y^2 \right)\hat{m} \right)^2}{\mathcal{G}} + 4iaA\left( aA\hat{\omega}+\hat{m} \right)sy  &= 0\,, \label{eq:gY}
\end{align}
\end{widetext}
where
\begin{align}
\hat{m}&\equiv \frac{m}{K_\phi\Gamma^2}\,,&
\hat{\omega}&=\frac{\omega}{\Ktau\Gamma^2}\,,
\end{align}
and the symbol $\mathcal{G}$ represents either $\mathcal{G}(x)$ or $\mathcal{G}(y)$; which one is clear from the context.

The equation for the Debye potential for scalar field ($S=0$) is in fact just a massless Klein\,--\,Gordon equation (complex conjugated). All the field components $\Phi^{(s,S)}_{[2s,0]}$ with $s=-S,\,-S+1,\,\dots,\,S-1,\,S$ are generated from the Debye potential $\hat{\psi}_{[0,2S]}$, see \cite{Keg1,Keg2} for neutrino field, Maxwell field and gravitational perturbations; for Rarita\,--\,Schwinger field see \cite{RS,RS2} (following equations can be proven to be equivalent with their results) and leads to 
\begin{align}
\Psi_0 = \Phi^{(2,2)}_{[4,0]} &= \left( \Th-\bar{\rho} \right)^3\left( \Th+3\bar{\rho} \right)\hat{\psi}_{[0,-4]}\,,\nonumber\\
\Sigma^{RS}_0 = \Phi^{(\nicefrac{3}{2},\nicefrac{3}{2})}_{[3,0]} &= \left( \Th-\bar{\rho} \right)^2\left( \Th+2\bar{\rho} \right)\hat{\psi}_{[0,-3]}\,,\nonumber\\
\Phi_0 = \Phi^{(1,1)}_{[2,0]} &= \left( \Th -\bar{\rho} \right)\left( \Th+\bar{\rho} \right)\hat{\psi}_{[0,-2]}\,, \nonumber\\
\chi_0 = \Phi^{(\nicefrac{1}{2},\nicefrac{1}{2})}_{[1,0]} &= \Th \hat{\psi}_{[0,-1]}\,, \nonumber\\
\Phi_{KG} = \Phi^{(0,0)}_{[0,0]}&= \hat{\psi}_{[0,0]}\,,\nonumber\\
\chi_1 = \Phi^{(-\nicefrac{1}{2},\nicefrac{1}{2})}_{[-1,0]} &= \Ed' \, \hat{\psi}_{[0,-1]}\,, \nonumber\\
\Phi_2 = \Phi^{(-1,1)}_{[-2,0]} &= \left( \Ed' -\bar{\tau} \right)\left( \Ed' +\bar{\tau} \right)\hat{\psi}_{[0,-2]}\,, \nonumber\\
\end{align}
\begin{align}
\Sigma^{RS}_3 = \Phi^{(-\nicefrac{3}{2},\nicefrac{3}{2})}_{[-3,0]} &= \left( \Ed' -\bar{\tau} \right)^2\left( \Ed' +2\bar{\tau} \right)\hat{\psi}_{[0,-3]}\,, \nonumber\\
\Psi_4 = \Phi^{(-2,2)}_{[-4,0]} &= \left( \Ed'-\bar{\tau} \right)^3\left( \Ed'+3\bar{\tau} \right)\hat{\psi}_{[0,-4]}\,,\nonumber
\end{align}
or, in general
\begin{align}
\Phi^{(S,S)}_{[2S,0]} &= \left( \Th-\bar{\rho} \right)^{2S-1}\left( \Th+(2S-1)\bar{\rho} \right)\hat{\psi}_{[0,-2S]} \\
\Phi^{(-S,S)}_{[-2S,0]} &= \left( \Ed'-\bar{\tau} \right)^{2S-1}\left( \Ed'+(2S-1)\bar{\tau} \right)\hat{\psi}_{[0,-2S]} 
\label{}
\end{align}
These equations can be beautifully simplified by application the following easy-to-prove identities
\begin{align}
\left( \Th-\bar{\rho} \right)\left( \Th+w\bar{\rho} \right) &= 
\left( \Th+(w-1)\bar{\rho} \right)\Th \,, \\
\left( \Ed'-\bar{\tau} \right)\left( \Ed'+w\bar{\tau} \right) &=
\left( \Ed'+(w-1)\bar{\tau} \right)\Ed'  \,,
\label{}
\end{align}
to the form
\begin{align}
\Phi^{(S,S)}_{[2S,0]} &= \Th^{2S} \hat{\psi}_{[0,-2S]}\,, 
& \Phi^{(-S,S)}_{[-2S,0]} &= \Ed^{\prime\, 2S} \hat{\psi}_{[0,-2S]} \,. 
\label{eq:ExS}
\end{align}

Note that operators $\Ed$ and $\Ed'$ incorporate the derivatives with respect to $\tau$ (multiplication by $i\omega$), $\phi$ (multiplication by $im$) and (the only nontrivial derivative) with respect to $x$, according to \re{PND-m}. 

Furthermore, the following symmetry holds
\begin{equation}
\left(\Phi^{(s,S)}_{[2s,0]}\right)^{\prime *} = i^{2S} \Phi^{(-s,S)}_{[-2s,0]}\,,
\label{eq:}
\end{equation}
which expresses the results as a derivative of the Debye potential $\hat{\psi}_{[0,-2S]}$ and interchanges $\Th \leftrightarrow \Ed'$ in \re{ExS}.

\section{Radial function}\label{sec:radial}

The \emph{regular} singular points of the equation \re{gY} are
\begin{equation}
-\frac{1}{A\Rm},\,-\frac{1}{A\Rp},\,-1,\,1,\,\infty\,.
\end{equation}

The behavior of the two linearly independent solutions around the outer black hole horizon is governed by characteristic exponents of the regular singular point $y=-1/A\Rp$:
\begin{align}
e_1 &= \phantom{-}\frac{S}{2} - i\frac{a}{\Rp-\Rm}\,\left( \hat{m}-\frac{\Ntau}{\Nphi\Kphi}\,\frac{\hat{\omega}}{\Omega_H}\right)\,, \\
e_2 &= -\frac{S}{2} + i\frac{a}{\Rp-\Rm}\,\left( \hat{m}-\frac{\Ntau}{\Nphi\Kphi}\,\frac{\hat{\omega}}{\Omega_H}\right)\,,
\label{}
\end{align}
where the angular velocity of the outer horizon $\Omega_H$ is defined with respect to the Killing vector field $\vec{\xi_H} = \vec{\xi_\tau}+\Omega_H \vec{\xi_\phi}$ and reads
\begin{equation}
\Omega_H = -\frac{a}{A}\,\frac{\Ktau \Ntau}{\Kphi\Nphi}\, \frac{a^2+\Rp^2}{1-A^2\Rp^2}\,.
\label{eq:AngVel}
\end{equation}
Then the solutions behave as
\begin{align}
\mathcal{Y}_1 & \sim \left(1+A\Rp y\right)^{e_1}\left( c_0 + c_1\left( 1+A\Rp y \right)+\dots \right),\\
\mathcal{Y}_2 & \sim \left(1+A\Rp y\right)^{e_2}\left( d_0 + d_1\left( 1+A\Rp y \right)+\dots \right) \,,
\label{}
\end{align}
if $e_1-e_2 \notin \mathbb{N}$, otherwise logarithmic terms will appear in $\mathcal{Y}_2$.

In the extremal case ($\Rp =\Rm$) the point $y=-1/A\Rp$ is in general \emph{irregular} singular point, and, thus, the asymptotic behaviour is difficult to investigate. But, in the case of static and axisymmetric configurations $y=-1/A\Rp$ becomes \emph{regular} singular point, and the exponents then depend on the eigenvalues $\Lambda^{(S)}_{(l0)}$, which we found in \cite{KofSep1} to be
\begin{equation}
\Lambda^{(S)}_{(l0)} = \left( 1-A^2\Rp^2 \right)\left[l\left( l+1 \right) + \frac{1}{3}\left( 1-S^2 \right)\right],
\end{equation}
and the calculations lead to the exponents\footnote{In \cite{KofSep1} the exponents were $(l-s,\,-l-s-1)$, but our current definition of the radial function $\mathcal{Y}$ differs from the one used in \cite{KofSep1} by the factor $\mathcal{G}^{s/2}(y)$.}
\begin{align}
e_1 & = l\,, & e_2 &= -l-1\,.
\label{eq:exponents}
\end{align}
In this extremal, static and axisymmetric case the Eq. \re{gY} can be solved exactly, one of the solution is simply polynomial, the second one is polynomial containing logarithms, see \cite{KofSep1}, where we also found the regular solution of the angular equation.

\section{Electromagnetic field --- the Meissner effect}\label{sec:elmag}
By reconstructing the self conjugated electromagnetic tensor $\vec{F}^*=\vec{F}+\frac{i}{2}\,\vec{\star F}$ from NP field components and in the null tetrad \re{PND-l}\,--\,\re{PND-m} we obtain 
\begin{multline}
F^*_{x\phi} = \Omega_0^2 \Kphi \Biggl[ i\left( 1+a^2A^2y^2 \right)\Phi_1\\ + \frac{a}{2}\,\frac{1-x^2}{\sqrt{\mathcal{G}(x)}}\,\sqrt{-\varepsilon \mathcal{G}(y)} \left( -\varepsilon^{-1} \Phi_0 + \Phi_2 \right) \Biggr]\,,
\label{eq:Flux}
\end{multline}
which represents the density of flux of ``magnetic + i $\times$ electric'' field.

The last NP component of electromagnetic field $\Phi_1$ was given in \cite{Keg1} as
\begin{equation}
2\Phi_1 
  = \Bigl[ \left( \Th+\rho-\bar{\rho} \right)\left( \Ed'+\bar{\tau} \right) 
  +\left( \Ed'+\tau'-\bar{\tau} \right)\left( \Th+\bar{\rho} \right) \Bigr]\hat{\psi}_{[0,-2]}
\nonumber 
\end{equation}
which can be again simplified. Thus, all the NP components of electromagnetic field in terms of the Debye potential $\hat{\psi}_{[0,-2]}$ read
\begin{align}
\Phi_0  & = \Th\Th\,\hat{\psi}_{[0,-2]} \label{eq:P0}\\
2\Phi_1 
 & = \left[ \left( \Th+\rho \right)\Ed' +\left( \Ed'+\tau' \right)\Th \right]\hat{\psi}_{[0,-2]}\\
\Phi_2 & =\Ed'\Ed'\hat{\psi}_{[0,-2]} \,. \label{eq:P2}
\end{align}

Meissner effect affect static axially symmetric field configurations in the vicinity of rapidly rotating outer horizon. For the extremal case the radial equation can be solved explicitly as a rational function, see \cite{KofSep1}, but only the series expansion above the horizon is necessary\footnote{Here we denote the solution by all of the separation constant: $l$, $m=0$ and spin $S=1$ as $Y^{(1)}_{(1l0)}$.}:
\begin{equation}
Y^{(1)}_{(1l0)} \sim \left( 1+A\Rp y \right)^l\left( c_0+c_1\left( 1+A\Rp y \right)^2\dots \right)\,.
\label{eq:DebSol}
\end{equation}
Calculating the field components $\Phi_0$, $\Phi_1$ and $\Phi_2$ using \re{P0}\,--\,\re{P2} and evaluating the density of the flux \re{Flux} in the limit $y\rightarrow -1/A\Rp$ with the solution \re{DebSol} yields
\begin{equation}
\lim_{y\rightarrow -1/A\Rp} F^*_{x\phi} = 0\,,
\label{eq:}\end{equation}
independently of the exact knowledge of the angular function $\mathcal{X}^{(1)}_{(1l0)}(x)$.

This means that there is no flux of static axially symmetric magnetic or electric field through the outer black hole horizon of extremely rotating uniformly accelerated black hole.

\section{Conclusions}
We have separated the Teukolsky master equation on a \emph{rotating C\,--\,metric} background for a test field of arbitrary spin $S$. We have separated also the equation for the Debye potential of these fields.

Utilizing the Debye potential for electromagnetic field we calculated the component $F^*_{x\phi}$, which represents the density of flux through surfaces $y=\;$const and $\tau=\;$const, of a electromagnetic field tensor and we have showed that for extremely rotating C\,--\,metric this flux is zero in the case of axially symmetric field configurations.

We have reformulated some results on the Debye potential in GHP formalism by applying a unifying notation and by using commutation relations simplifying known results.

\begin{acknowledgements}
D.K. acknowledges the support from the Czech Science Foundation, Grant No. 14-37086G --- the Albert Einstein Centre. Moreover, D.K. would like to thank Prof. J. Bi\v{c}\'ak for introducing him to the C\,--\,metric, to Dr. M. Scholtz  and Dr. Georgios Loukes-Gerakopoulos for inspiring discussions and comments on the manuscript.

We were unaware of the previous work of T. Prestidge \cite{Prestidge1998} and D. Bini, C. Cherubini and A. Geralico \cite{Bini2008} and therefore we would like to apologize for not paying the proper attention to their pioneering work in \cite{KofSep1}.
\end{acknowledgements}


\begin{thebibliography}{21}%
\makeatletter
\providecommand \@ifxundefined [1]{%
 \@ifx{#1\undefined}
}%
\providecommand \@ifnum [1]{%
 \ifnum #1\expandafter \@firstoftwo
 \else \expandafter \@secondoftwo
 \fi
}%
\providecommand \@ifx [1]{%
 \ifx #1\expandafter \@firstoftwo
 \else \expandafter \@secondoftwo
 \fi
}%
\providecommand \natexlab [1]{#1}%
\providecommand \enquote  [1]{``#1''}%
\providecommand \bibnamefont  [1]{#1}%
\providecommand \bibfnamefont [1]{#1}%
\providecommand \citenamefont [1]{#1}%
\providecommand \href@noop [0]{\@secondoftwo}%
\providecommand \href [0]{\begingroup \@sanitize@url \@href}%
\providecommand \@href[1]{\@@startlink{#1}\@@href}%
\providecommand \@@href[1]{\endgroup#1\@@endlink}%
\providecommand \@sanitize@url [0]{\catcode `\\12\catcode `\$12\catcode
  `\&12\catcode `\#12\catcode `\^12\catcode `\_12\catcode `\%12\relax}%
\providecommand \@@startlink[1]{}%
\providecommand \@@endlink[0]{}%
\providecommand \url  [0]{\begingroup\@sanitize@url \@url }%
\providecommand \@url [1]{\endgroup\@href {#1}{\urlprefix }}%
\providecommand \urlprefix  [0]{URL }%
\providecommand \Eprint [0]{\href }%
\providecommand \doibase [0]{http://dx.doi.org/}%
\providecommand \selectlanguage [0]{\@gobble}%
\providecommand \bibinfo  [0]{\@secondoftwo}%
\providecommand \bibfield  [0]{\@secondoftwo}%
\providecommand \translation [1]{[#1]}%
\providecommand \BibitemOpen [0]{}%
\providecommand \bibitemStop [0]{}%
\providecommand \bibitemNoStop [0]{.\EOS\space}%
\providecommand \EOS [0]{\spacefactor3000\relax}%
\providecommand \BibitemShut  [1]{\csname bibitem#1\endcsname}%
\let\auto@bib@innerbib\@empty
\bibitem [{\citenamefont {Prestidge}(1998)}]{Prestidge1998}%
  \BibitemOpen
  \bibfield  {author} {\bibinfo {author} {\bibfnamefont {T.}~\bibnamefont
  {Prestidge}},\ }\href {\doibase 10.1103/PhysRevD.58.124022} {\bibfield
  {journal} {\bibinfo  {journal} {Phys. Rev. D}\ }\textbf {\bibinfo {volume}
  {58}},\ \bibinfo {pages} {124022} (\bibinfo {year} {1998})}\BibitemShut
  {NoStop}%
\bibitem [{\citenamefont {Bini}\ \emph {et~al.}(2008)\citenamefont {Bini},
  \citenamefont {Cherubini},\ and\ \citenamefont {Geralico}}]{Bini2008}%
  \BibitemOpen
  \bibfield  {author} {\bibinfo {author} {\bibfnamefont {D.}~\bibnamefont
  {Bini}}, \bibinfo {author} {\bibfnamefont {C.}~\bibnamefont {Cherubini}}, \
  and\ \bibinfo {author} {\bibfnamefont {A.}~\bibnamefont {Geralico}},\ }\href
  {\doibase http://dx.doi.org/10.1063/1.2938699} {\bibfield  {journal}
  {\bibinfo  {journal} {Journal of Mathematical Physics}\ }\textbf {\bibinfo
  {volume} {49}},\ \bibinfo {eid} {062502} (\bibinfo {year} {2008}),\
  http://dx.doi.org/10.1063/1.2938699}\BibitemShut {NoStop}%
\bibitem [{\citenamefont {Kofro\v{n}}(2015)}]{KofSep1}%
  \BibitemOpen
  \bibfield  {author} {\bibinfo {author} {\bibfnamefont {D.}~\bibnamefont
  {Kofro\v{n}}},\ }\href {\doibase 10.1103/PhysRevD.92.124064} {\bibfield
  {journal} {\bibinfo  {journal} {Phys. Rev. D}\ }\textbf {\bibinfo {volume}
  {92}},\ \bibinfo {pages} {124064} (\bibinfo {year} {2015})}\BibitemShut
  {NoStop}%
\bibitem [{\citenamefont {{Hong}}\ and\ \citenamefont {{Teo}}(2005)}]{ht2}%
  \BibitemOpen
  \bibfield  {author} {\bibinfo {author} {\bibfnamefont {K.}~\bibnamefont
  {{Hong}}}\ and\ \bibinfo {author} {\bibfnamefont {E.}~\bibnamefont {{Teo}}},\
  }\href@noop {} {\bibfield  {journal} {\bibinfo  {journal} {Class. and Quantum
  Grav.}\ }\textbf {\bibinfo {volume} {22}},\ \bibinfo {pages} {109} (\bibinfo
  {year} {2005})}\BibitemShut {NoStop}%
\bibitem [{\citenamefont {Bi\v{c}{\'{a}}k}\ and\ \citenamefont
  {Kofro\v{n}}(2009)}]{BiKofAcc}%
  \BibitemOpen
  \bibfield  {author} {\bibinfo {author} {\bibfnamefont {J.}~\bibnamefont
  {Bi\v{c}{\'{a}}k}}\ and\ \bibinfo {author} {\bibfnamefont {D.}~\bibnamefont
  {Kofro\v{n}}},\ }\href@noop {} {\bibfield  {journal} {\bibinfo  {journal}
  {Gen. Relativ. Gravit.}\ }\textbf {\bibinfo {volume} {41}},\ \bibinfo {pages}
  {1981} (\bibinfo {year} {2009})}\BibitemShut {NoStop}%
\bibitem [{\citenamefont {Penna}(2014)}]{Penna}%
  \BibitemOpen
  \bibfield  {author} {\bibinfo {author} {\bibfnamefont {R.~F.}\ \bibnamefont
  {Penna}},\ }\href {\doibase 10.1103/PhysRevD.90.043003} {\bibfield  {journal}
  {\bibinfo  {journal} {Phys. Rev. D}\ }\textbf {\bibinfo {volume} {90}},\
  \bibinfo {pages} {043003} (\bibinfo {year} {2014})}\BibitemShut {NoStop}%
\bibitem [{\citenamefont {Bi\v{c}\'{a}k}\ and\ \citenamefont
  {Jani\v{s}}(1985)}]{BicakJanis}%
  \BibitemOpen
  \bibfield  {author} {\bibinfo {author} {\bibfnamefont {J.}~\bibnamefont
  {Bi\v{c}\'{a}k}}\ and\ \bibinfo {author} {\bibfnamefont {V.}~\bibnamefont
  {Jani\v{s}}},\ }\href {\doibase 10.1093/mnras/212.4.899} {\bibfield
  {journal} {\bibinfo  {journal} {MNRAS}\ }\textbf {\bibinfo {volume} {212}},\
  \bibinfo {pages} {899} (\bibinfo {year} {1985})}\BibitemShut {NoStop}%
\bibitem [{\citenamefont {{Dov\v{c}iak}}\ \emph {et~al.}(2000)\citenamefont
  {{Dov\v{c}iak}}, \citenamefont {{Karas}},\ and\ \citenamefont
  {{Lanza}}}]{DKL}%
  \BibitemOpen
  \bibfield  {author} {\bibinfo {author} {\bibfnamefont {M.}~\bibnamefont
  {{Dov\v{c}iak}}}, \bibinfo {author} {\bibfnamefont {V.}~\bibnamefont
  {{Karas}}}, \ and\ \bibinfo {author} {\bibfnamefont {A.}~\bibnamefont
  {{Lanza}}},\ }\href {\doibase 10.1088/0143-0807/21/4/304} {\bibfield
  {journal} {\bibinfo  {journal} {European Journal of Physics}\ }\textbf
  {\bibinfo {volume} {21}},\ \bibinfo {pages} {303} (\bibinfo {year}
  {2000})}\BibitemShut {NoStop}%
\bibitem [{\citenamefont {{Bi{\v c}{\'a}k}}\ and\ \citenamefont
  {{Ledvinka}}(2000)}]{BicakLedvinka}%
  \BibitemOpen
  \bibfield  {author} {\bibinfo {author} {\bibfnamefont {J.}~\bibnamefont
  {{Bi{\v c}{\'a}k}}}\ and\ \bibinfo {author} {\bibfnamefont {T.}~\bibnamefont
  {{Ledvinka}}},\ }\href@noop {} {\bibfield  {journal} {\bibinfo  {journal}
  {Nuovo Cimento B Serie}\ }\textbf {\bibinfo {volume} {115}},\ \bibinfo
  {pages} {739} (\bibinfo {year} {2000})},\ \Eprint
  {http://arxiv.org/abs/gr-qc/0012006} {gr-qc/0012006} \BibitemShut {NoStop}%
\bibitem [{\citenamefont {Karas}\ and\ \citenamefont
  {Budinová}(2000)}]{KarasBudinova}%
  \BibitemOpen
  \bibfield  {author} {\bibinfo {author} {\bibfnamefont {V.}~\bibnamefont
  {Karas}}\ and\ \bibinfo {author} {\bibfnamefont {Z.}~\bibnamefont
  {Budinová}},\ }\href {http://stacks.iop.org/1402-4896/61/i=2/a=014}
  {\bibfield  {journal} {\bibinfo  {journal} {Physica Scripta}\ }\textbf
  {\bibinfo {volume} {61}},\ \bibinfo {pages} {253} (\bibinfo {year}
  {2000})}\BibitemShut {NoStop}%
\bibitem [{\citenamefont {{Pleba{\'{n}}ski}}\ and\ \citenamefont
  {{Demia{\'{n}}ski}}(1976)}]{pd}%
  \BibitemOpen
  \bibfield  {author} {\bibinfo {author} {\bibfnamefont {J.~F.}\ \bibnamefont
  {{Pleba{\'{n}}ski}}}\ and\ \bibinfo {author} {\bibfnamefont {M.}~\bibnamefont
  {{Demia{\'{n}}ski}}},\ }\href@noop {} {\bibfield  {journal} {\bibinfo
  {journal} {Annals of Physics}\ }\textbf {\bibinfo {volume} {98}},\ \bibinfo
  {pages} {98} (\bibinfo {year} {1976})}\BibitemShut {NoStop}%
\bibitem [{\citenamefont {{Teukolsky}}(1973)}]{Teuk}%
  \BibitemOpen
  \bibfield  {author} {\bibinfo {author} {\bibfnamefont {S.~A.}\ \bibnamefont
  {{Teukolsky}}},\ }\href {\doibase 10.1086/152444} {\bibfield  {journal}
  {\bibinfo  {journal} {Astrophys. J.}\ }\textbf {\bibinfo {volume} {185}},\
  \bibinfo {pages} {635} (\bibinfo {year} {1973})}\BibitemShut {NoStop}%
\bibitem [{\citenamefont {del Castillo}(1989)}]{RS}%
  \BibitemOpen
  \bibfield  {author} {\bibinfo {author} {\bibfnamefont {G.~F.~T.}\
  \bibnamefont {del Castillo}},\ }\href@noop {} {\bibfield  {journal} {\bibinfo
   {journal} {J. Math. Phys.}\ }\textbf {\bibinfo {volume} {30}},\ \bibinfo
  {pages} {446} (\bibinfo {year} {1989})}\BibitemShut {NoStop}%
\bibitem [{\citenamefont {{Cohen}}\ and\ \citenamefont
  {{Kegeles}}(1975)}]{Keg1}%
  \BibitemOpen
  \bibfield  {author} {\bibinfo {author} {\bibfnamefont {J.~M.}\ \bibnamefont
  {{Cohen}}}\ and\ \bibinfo {author} {\bibfnamefont {L.~S.}\ \bibnamefont
  {{Kegeles}}},\ }\href {\doibase 10.1016/0375-9601(75)90583-6} {\bibfield
  {journal} {\bibinfo  {journal} {Physics Letters A}\ }\textbf {\bibinfo
  {volume} {54}},\ \bibinfo {pages} {5} (\bibinfo {year} {1975})}\BibitemShut
  {NoStop}%
\bibitem [{\citenamefont {{Kegeles}}\ and\ \citenamefont
  {{Cohen}}(1979)}]{Keg2}%
  \BibitemOpen
  \bibfield  {author} {\bibinfo {author} {\bibfnamefont {L.~S.}\ \bibnamefont
  {{Kegeles}}}\ and\ \bibinfo {author} {\bibfnamefont {J.~M.}\ \bibnamefont
  {{Cohen}}},\ }\href {\doibase 10.1103/PhysRevD.19.1641} {\bibfield  {journal}
  {\bibinfo  {journal} {Phys. Rev.~D}\ }\textbf {\bibinfo {volume} {19}},\
  \bibinfo {pages} {1641} (\bibinfo {year} {1979})}\BibitemShut {NoStop}%
\bibitem [{\citenamefont {{Geroch}}\ \emph {et~al.}(1973)\citenamefont
  {{Geroch}}, \citenamefont {{Held}},\ and\ \citenamefont {{Penrose}}}]{GHP}%
  \BibitemOpen
  \bibfield  {author} {\bibinfo {author} {\bibfnamefont {R.}~\bibnamefont
  {{Geroch}}}, \bibinfo {author} {\bibfnamefont {A.}~\bibnamefont {{Held}}}, \
  and\ \bibinfo {author} {\bibfnamefont {R.}~\bibnamefont {{Penrose}}},\ }\href
  {\doibase 10.1063/1.1666410} {\bibfield  {journal} {\bibinfo  {journal}
  {Journal of Mathematical Physics}\ }\textbf {\bibinfo {volume} {14}},\
  \bibinfo {pages} {874} (\bibinfo {year} {1973})}\BibitemShut {NoStop}%
\bibitem [{\citenamefont {{Penrose}}\ and\ \citenamefont
  {{Rindler}}(1984)}]{RindlerPenrose1}%
  \BibitemOpen
  \bibfield  {author} {\bibinfo {author} {\bibfnamefont {R.}~\bibnamefont
  {{Penrose}}}\ and\ \bibinfo {author} {\bibfnamefont {W.}~\bibnamefont
  {{Rindler}}},\ }\href@noop {} {\emph {\bibinfo {title} {Spinors and
  space-time.~Vol.~1: Two-spinor calculus and relativistic fields..~R.~Penrose,
  W.~Rindler.Cambridge University Press, Cambridge - London - New York - New
  Rochelle - Melbourne - Sydney.~458 pp.~Price {\pounds} 45.00, \$ 89.50
  (1984).}}}\ (\bibinfo {year} {1984})\BibitemShut {NoStop}%
\bibitem [{\citenamefont {Kinnersley}\ and\ \citenamefont {Walker}(1970)}]{kw}%
  \BibitemOpen
  \bibfield  {author} {\bibinfo {author} {\bibfnamefont {W.}~\bibnamefont
  {Kinnersley}}\ and\ \bibinfo {author} {\bibfnamefont {M.}~\bibnamefont
  {Walker}},\ }\href@noop {} {\bibfield  {journal} {\bibinfo  {journal} {Phys.
  Rev.~D}\ }\textbf {\bibinfo {volume} {2}},\ \bibinfo {pages} {1359} (\bibinfo
  {year} {1970})}\BibitemShut {NoStop}%
\bibitem [{\citenamefont {{Griffiths}}\ \emph {et~al.}(2006)\citenamefont
  {{Griffiths}}, \citenamefont {{Krtou\v{s}}},\ and\ \citenamefont
  {{Podolsk{\'y}}}}]{gkp}%
  \BibitemOpen
  \bibfield  {author} {\bibinfo {author} {\bibfnamefont {J.~B.}\ \bibnamefont
  {{Griffiths}}}, \bibinfo {author} {\bibfnamefont {P.}~\bibnamefont
  {{Krtou\v{s}}}}, \ and\ \bibinfo {author} {\bibfnamefont {J.}~\bibnamefont
  {{Podolsk{\'y}}}},\ }\href@noop {} {\bibfield  {journal} {\bibinfo  {journal}
  {Class. and Quantum Grav.}\ }\textbf {\bibinfo {volume} {23}},\ \bibinfo
  {pages} {6745} (\bibinfo {year} {2006})}\BibitemShut {NoStop}%
\bibitem [{\citenamefont {{Kubiz{\v n}{\'a}k}}\ and\ \citenamefont {{Krtou{\v
  s}}}(2007)}]{KrKuCKYPD}%
  \BibitemOpen
  \bibfield  {author} {\bibinfo {author} {\bibfnamefont {D.}~\bibnamefont
  {{Kubiz{\v n}{\'a}k}}}\ and\ \bibinfo {author} {\bibfnamefont
  {P.}~\bibnamefont {{Krtou{\v s}}}},\ }\href {\doibase
  10.1103/PhysRevD.76.084036} {\bibfield  {journal} {\bibinfo  {journal} {Phys.
  Rev.~D}\ }\textbf {\bibinfo {volume} {76}},\ \bibinfo {eid} {084036}
  (\bibinfo {year} {2007})},\ \Eprint {http://arxiv.org/abs/0707.0409}
  {arXiv:0707.0409 [gr-qc]} \BibitemShut {NoStop}%
\bibitem [{\citenamefont {Torres~del Castillo}(1989)}]{RS2}%
  \BibitemOpen
  \bibfield  {author} {\bibinfo {author} {\bibfnamefont {G.~F.}\ \bibnamefont
  {Torres~del Castillo}},\ }\href {\doibase http://dx.doi.org/10.1063/1.528312}
  {\bibfield  {journal} {\bibinfo  {journal} {Journal of Mathematical Physics}\
  }\textbf {\bibinfo {volume} {30}},\ \bibinfo {pages} {1323} (\bibinfo {year}
  {1989})}\BibitemShut {NoStop}%
\end{thebibliography}
%

\end{document}